\documentclass[twocolumn,showpacs,preprintnumbers,amsmath,amssymb]{revtex4}

\usepackage{graphicx}
\usepackage{dcolumn}
\usepackage{bm}

\begin{document}

\title{Anomalies in the superconducting dome of the Bi-cuprates}

\author{L. Dudy$^{1,2}$}
\author{A. Krapf$^2$}
\author{H. Dwelk}
\author{S. Rogaschewski$^2$}
\author{B. M\"uller$^2$}
\author{O. L\"ubben$^2$}
\author{C. Janowitz$^2$}
\author{R. Manzke$^2$}

\affiliation{$^1$Randall Laboratory, University of Michigan, Ann
Arbor, MI 48109-1040, USA \\ $^2$Humboldt-Universit\"at zu Berlin,
Institut f\"ur Physik, {Newtonstr{.15}}, D-12489 Berlin, Germany}%

\date{\today}

\begin{abstract}
We report characterization results by energy dispersive x-ray
analysis and AC-susceptibility for a statistically relevant number
of Bi$_{2+z-y}$Pb$_y$Sr$_{2-x-z}$La$_x$CuO$_{6+\delta}$ single
crystals. We show that the two structurally quite different
modifications of the single-layered Bi-cuprate, namely
Bi$_{2+z-y}$Pb$_y$Sr$_{2-x-z}$La$_x$CuO$_{6+\delta}$   with y=0.4
and Bi$_{2+z}$Sr$_{2-x-z}$La$_x$CuO$_{6+\delta}$ , exhibit anomalies
in the superconducting transition temperature at certain hole
doping, e.g. at 1/8 holes per Cu. These doping values agree well
with the 'magic doping fractions' found in the temperature dependent
resistance of La$_{2-x}$Sr$_x$CuO$_4$ by Komiya et al.
\cite{Komiya2005}. This new set of findings suggests that all these
anomalies are generic for the hole-doped high-temperature
superconductors.
\end{abstract}

\maketitle

\section{Introduction}
One central paradigm in research of the hole-doped high-temperature
superconducting cuprates (HTSC's) is the so-called 'generic
phase-diagram', where the appearance of  phases is shown in relation
to the hole-doping and temperature.  In this generic phase-diagram
the superconducting dome, the superconducting transition temperature
(T$_C$) relative to the hole-doping, is typically illustrated as a
flipped parabola which exhibits the maximum at around 16 percent
holes per Cu atom. This parabola is called the 'universal curve'
\cite{Presland1991} and is suggested to be universal for all HTSC
materials. The differences in the various HTSC-materials are
accounted for by commonly scaling this parabola, in most cases just
by scaling to maximum T$_C$. In one very practical application, an
appropriately scaled parabola can then be used for straightforward
estimation of a samples' hole-doping. This estimation is often done
especially in the Bi-cuprates where the doping can hardly be
determined exactly.

Beside the practical aspect of having a simple and smooth parabola,
there is also the possibility that there are generic anomalies of
T$_C$ within the superconducting dome. There seems to be at least
one certain hole-doping value where T$_C$ shows an anomaly and is
suppressed.  Historically, the first anomaly found was in
La$_{2-x}$Ba$_x$CuO$_4$ (LBCO) at a hole-doping value of 1/8 by
Moodenbaugh et al. \cite{Moodenbaugh1988}. Also
La$_{2-x}$Sr$_x$CuO$_4$ (LSCO) has a 1/8 anomaly, but additionally,
other fractional anomalies (so called 'magic doping fractions') can
be suggested for this material \cite{Zhou2004,Komiya2005}.  Komiya
et al. \cite{Komiya2005} extracted for the hole-doping of these the
relation p=(2m+1)/2n, where p is the hole-doping and m and n are
integers. In order to complete the list of known anomalies,
YBa$_2$Cu$_3$O$_{7-x}$ (YBCO) shows also a reduction in T$_C$ which
was found quite early and is often referred to as 'the 60K plateau'
\cite{Cava1987}. This '60K plateau' can be associated with 1/8
hole-doping value \cite{Tallon1997}. There also exists a '90K
plateau' in YBCO. For the Bi-cuprates, there is less knowledge about
the existence of a 1/8 anomaly or others. There is only one report
for Bi$_{2+z}$Sr$_{2-x-z}$La$_x$CuO$_{6+\delta}$  by Yang et al.
\cite{Yang2000}. It might therefore be interesting to examine
members of the Bi-cuprates for these anomalies. In the case that a
Bi-cuprate also shows anomalies at the same doping values like the
other HTSC's, it can be suggested that the simple parabola is only a
zeroth order approximation and an included finer structure is
generic for the HTSC's.

A generic existence of anomalies may also rely on important and
interesting physics which prominently shows up in the numerous
studies in connection with the 1/8 problem. Essentially, the
experiments indicate spin and charge-order within the CuO$_2$-plane
\cite{Tranquada1995, Yamada1997, Zimmermann1998, Abbamonte2005}
resulting in more insulating behavior
\cite{Moodenbaugh1988,Komiya2005} accompanied by a large single
particle gap at the antinodes \cite{Valla2006} as well as the
suppression of Josephson coupling perpendicular to the planes
\cite{Tranquada2008}. From a theoretical point, among many models
proclaimed so far, there exist some models which account for some,
if not all, findings about the 1/8 anomaly. In the phenomenological
view, examples for these possibilities are, at least, the two
formulations of so-called pair-density waves
\cite{Chen2004,Berg2007,Berg2009}. For the microscopic picture,
these phenomenological formulations might be respectively broken
down to the existence of a Wigner-like-crystal of Cooper pairs
\cite{Chen2004} or the formation of stripes \cite{Berg2007,
Berg2009}. Another microscopic model which might be relevant here is
that of a Wigner-like crystal of holes in an antiferromagnetic
background \cite{KimHor2001, KimHor2006}.

All these mentioned models have in common that they not only aim to
explain superconductivity but also account for the existence of, at
least, a 1/8 anomaly. Thus, there might be insights to gain by
examining not only the superconductivity of the HTSC's but also its
absence. So it comes that the present manuscript aims to discuss
that members of the Bi-cuprate exhibits anomalies of T$_C$ at
certain hole-dopings, similar to the ones found by Komiya et al.
\cite{Komiya2005}.  Shown here are results from single crystals of
the one-layers Bi$_{2+z}$Sr$_{2-x-z}$La$_x$CuO$_{6+\delta}$
(La-Bi2201) and Bi$_{2+z-y}$Pb$_y$Sr$_{2-x-z}$La$_x$CuO$_{6+\delta}$
((Pb,La)-Bi2201). The process of hole-doping in the Bi-cuprates is
quite complex: In Bi$_{2+z}$Sr$_{2-z}$CuO$_{6+\delta}$ (Bi2201), the
hole doping is done by extra oxygen and the amount of Bi on Sr
positions, whereas in La-Bi2201 the hole doping can be controlled by
lanthanum substitution only. Additionally substituting Pb on Bi
positions, i.e. (Pb,La)-Bi2201, leads with high enough amount of Pb
to structurally more clean crystals. These crystals are dominantly
in the so-called '$\beta$-phase' (see, e.g., \cite{Luebben2010})
which means that these crystals are free of an incommensurate quasi
1x5 superstructure. Since Pb is an active dopant and the extra
oxygen ($\delta$) is known to be reduced by Pb substitution (see,
e.g., \cite{Zhang1991}), the same lanthanum content will produce
different hole-dopings in both systems (see \cite{Ariffin2009}).

The organization of this manuscript is as follows. First, we will
show by a statistically relevant amount of characterization data
that in both of these systems, La-Bi2201 and (Pb,La)-Bi2201,
anomalies of T$_C$ exist at certain, but different, La
concentrations. This will be done by applying a refinement scheme
that first uses the full data obtained from 299 crystals and then
reduces the number of samples by defining restrictions, which are
meant to optimize the superconducting properties or 'cleanness' of
these materials. As both do show anomalies but at different La
content, this already suggests that the anomalies found cannot be a
feature of the La substitution but only depend on the hole doping
value. For LSCO, the existence of anomalies will be shown from a
collection of data from the literature. In LSCO, under correct
annealing conditions, the hole concentration is merely directly
proportional to the strontium substitution level. Therefore, the La
concentrations of the anomalies in La-Bi2201 and (Pb,La)-Bi2201 can
be assigned to a hole-doping value according to the hole
concentrations of the anomalies in LSCO. Eventually, a scaling
between the La- (and implicitly the Pb-) content relative to the
hole-doping is constructed. This hole-lanthanum scaling equals
perfectly the scaling obtained by x-ray absorption spectroscopy
(XAS) \cite{Schneider2005, Ariffin2009}. Because the opposite
direction can be argued, it is therefore shown that also in the two
single-layer Bi-cuprates anomalies at certain hole dopings exist.
This suggests that all HTSC's exhibit anomalies at always the same
hole dopings.

\section{Superconductivity in relation to the Lanthanum and Lead content}

\begin{figure*}
  \includegraphics[width=0.9\textwidth]{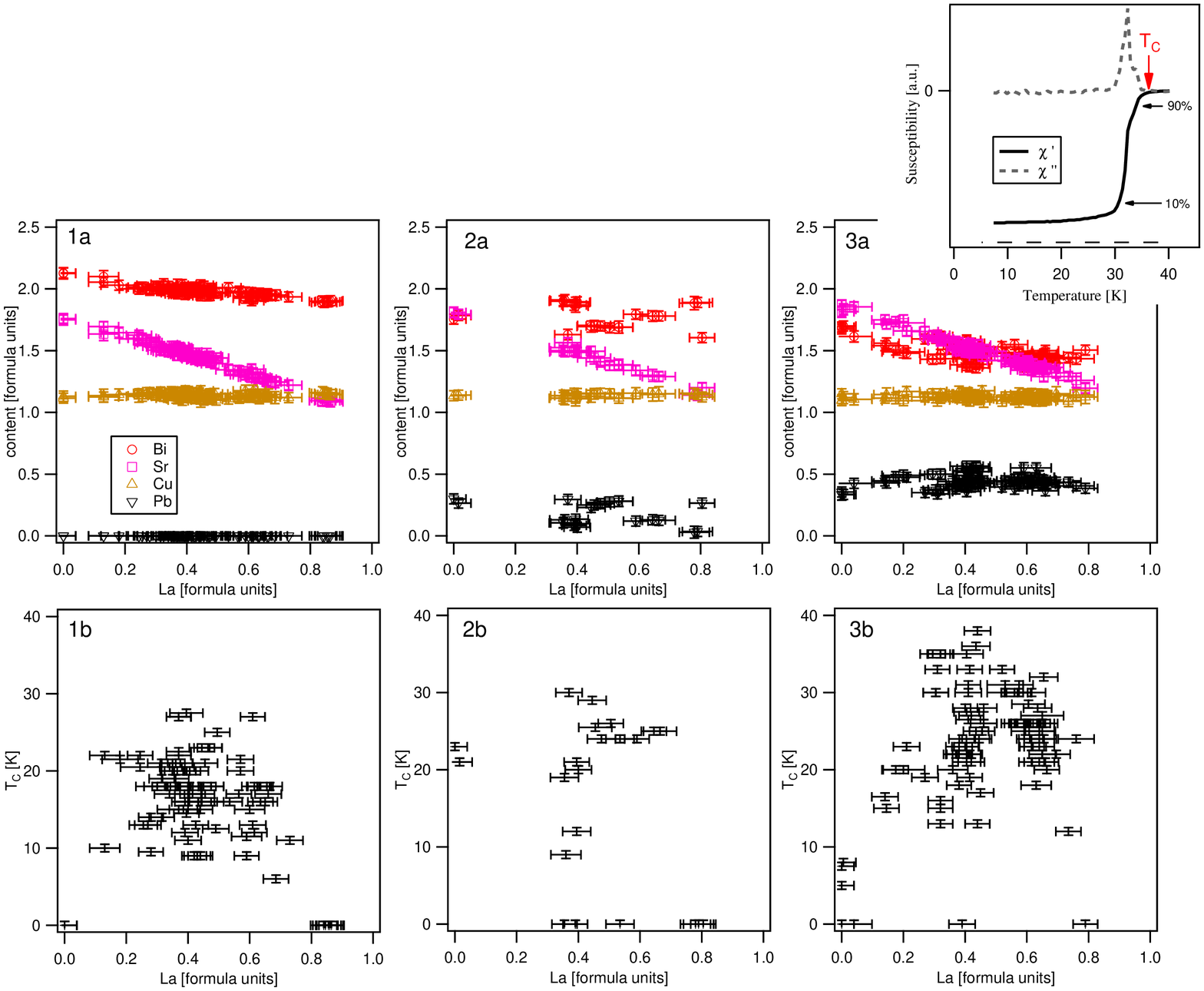}
  \caption{Characterization results of the
Bi$_{2+z-y}$Pb$_y$Sr$_{2-x-z}$La$_x$CuO$_{6+\delta}$ single crystals
as obtained by energy dispersive x-ray analysis (EDX) and
AC-susceptibility. The inset in the upper right shows a typical
AC-susceptibility result and the definitions used: The
superconducting transition temperature is the onset-T$_C$, the
transition width $\Delta$T$_C$ is the temperature range where the
real part of the susceptibility drops from 90\% to 10\% from the
paramagnetic state into the diamagnetic state. The top panels 1a, 2a
and 3a display the results of the EDX characterization. Shown is the
average chemical composition of Bi, Sr and Cu vs the average
lanthanum content. Panel 1a shows this for Pb-free samples (y=0),
whereas Panel 2a shows this for the low Pb-content of y$\in$]0;0.3]
and Panel 3a for the high Pb-content of y$\in$]0.3;0.6]. The lower
panels show the T$_C$ as obtained by AC-susceptibility vs the
average lanthanum content x. Panel 2a shows this for y=0, Panel 2b
for y$\in$]0;0.3] and Panel 3b for y$\in$]0.3;0.6].}\label{FIG1}
\end{figure*}

Single crystals of (Pb,La)-Bi2201 were grown by the self-flux
method; details were reported elsewhere \cite{Luebben2010}. The
samples were characterized by energy dispersive x-ray  analysis
(EDX) and AC-susceptibility. EDX probes the chemical composition for
each crystal and is computed standard-free by the use of the
PUzaf-correction \cite{Heckel1984}. To compute the elementary
composition for Cu, the K doublet was used. For La and Sr, the
L-series was used and for Bi and Pb the M series. By
AC-susceptibility, we obtain $\chi'$ and $\chi''$ which are
respectively the real and imaginary parts of the AC-susceptibility
times the volume of the sample and a filling factor. For the
measurements, a commercial Quantum Design PPMS 6000 system with 0.1
Oe as the amplitude of the magnetic field was used. The
AC-susceptibility for a (Pb,La)-Bi2201 single crystal is presented
in the inset of FIG. \ref{FIG1}. The real part $\chi'$ indicates an
onset-T$_C$ of 36 K and a transition range ($\Delta$T$_C$) of 3 K.
The transition range is derived from the temperature range using the
criterion that the real part changes in a 90 to 10 percent range
from the paramagnetic to the superconducting state.

The whole dataset contained 299 crystals which were optically
selected for smoothness and regular shape. Superconducting samples
were chosen with the restriction that only one peak in the imaginary
part of their AC-susceptibility is present. Initially, no
distinction was made between a small or broad transition width as
long as the peak in the imaginary part is Gaussian-like and the real
part downturn is smooth.

The chemical composition obtained by EDX was then fitted by a linear
dependence. For this, the Pb content (y) and the La content (x) were
assumed to be independent of each other. If the content of Bi, Sr
and Cu is written as c(Bi), c(Sr) and c(Cu), the linear dependencies
of these 299 samples are given by
\begin{eqnarray}
c(Bi)&=(2.08\pm0.01)-(0.21\pm0.01)x-(1.12\pm0.01)y,\nonumber\\
  & \chi^2=0.70246; \nonumber\\
c(Sr)&=(1.79\pm0.01)-(0.78\pm0.01)x+(0.12\pm0.01)y,\nonumber\\
& \chi^2=0.55005; \nonumber\\
 c(Cu)&=(1.13\pm0.01)-(0.01\pm0.01)x+(0.01\pm0.01)y,\nonumber\\
& \chi^2=0.60479. \nonumber
\end{eqnarray}
In order to define a clean phase, we apply a refinement scheme which
we call here the 'compositional restriction'.  For this restriction,
crystals with a deviation of the Bi, Sr or Cu composition from the
linear dependencies larger than 0.05 formula units were removed.  By
applying this restriction, the size of the dataset was reduced to
208 samples.

In the upper (a) panels of FIG. \ref{FIG1}, the chemical composition
of Bi, Sr and Cu versus the lanthanum content is shown. The
resulting T$_C$ relative to the Lanthanum content for the
compositional restricted dataset is shown in the lower (b) panels of
FIG. \ref{FIG1}. There, the data are shown in slices of the Pb
content, for y=0, y$\in$]0;0.3], and y$\in$]0.3;0.6]. It is clearly
indicated that the maximum achievable superconducting transition
temperature T$_C^{max}$ increases by increasing the amount of Pb.
Typically, the curve T$_C$ vs La content is described as a parabolic
curve (see, e.g., \cite{Ando2000}), but here the data points appear
strongly scattered or irregular although the dependence of Bi, Sr,
and Cu vs the lanthanum content looks smooth. For Pb concentrations
higher than y=0, this can be best observed when looking at the Bi
curve and the Pb curve, which have to be complementary in their
gradients and curvatures when Pb substitutes for Bi. For y=0, a
strong scattering of T$_C$ values takes place in the region around
x=0.25, 0.4, and 0.6 formula units. At y$\in$]0;0.3], no definitive
values can be found because the statistics are quite poor. It is
suggested that there is a center of scattering around x=0.4 formula
units. For y$\in$]0.3;0.6], a strong scattering is visible in the
region of x=0.4 and 0.65 formula units. We will see that this
scattering is not accidentally but that these regions contain
anomalies.

\subsection{Anomalies in T$_C$}

\begin{figure*}
  \includegraphics[width=1\textwidth]{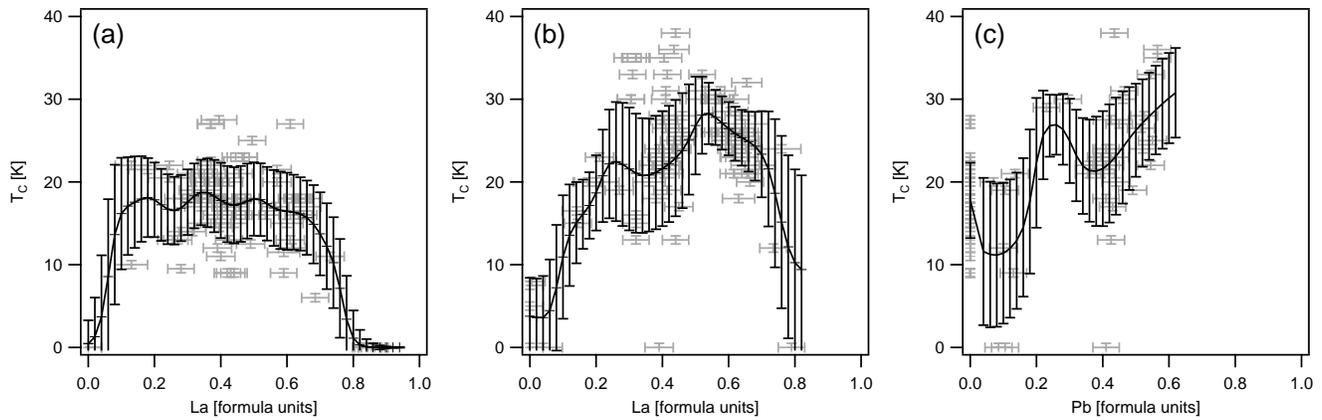}
  \caption{Averaging the T$_C$(x,y) values of the
Bi$_{2+z-y}$Pb$_y$Sr$_{2-x-z}$La$_x$CuO$_{6+\delta}$  single
crystals. The averaging algorithm relies on a Gaussian probability
distribution (see APPENDIX \ref{gaussianslotting}). The continuous
curve in all three panels is the averaged T$_C$(x,y), plotted here
for y=0 (a),  y=0.4 (b) and x=0.4 (c).  Also indicated is the
standard deviation for the averaging. The light gray crosses are the
original data-points T$_C$(x,y) as also shown in
FIG.\ref{FIG1}.}\label{FIG2}
\end{figure*}

The large number of data points makes it possible to apply some
statistical routines to gain a deeper insight into the true
three-dimensional T$_C$ vs La and Pb phase diagram. First, one can
average over all crystals of particular composition, which may act
to remove effects of a variation of the extra oxygen $\delta$ and
other experimental uncertainties. The averaging algorithm used and
described in the APPENDIX \ref{gaussianslotting}, relies on a
Gaussian probability distribution and includes all experimental
uncertainties. FIG. \ref{FIG2} shows the averages from that dataset
which was before refined by the compositional restriction. Three
selected cuts are shown given by the restrictions y=0 (a), y=0.4
(b), and x=0.4 (c), respectively. Also shown in this figure are the
individual datapoints which were used to compute this average. These
are marked by light gray crosses. The average T$_C$ curve for fixed
La upon the change in the Pb substitution, as shown in FIG.
\ref{FIG2} (c), indicates clearly an overall trend of an increase in
T$_C$ by increasing the amount of Pb. It is suggested that the
irregularity between 0 and 0.3 formula units of lead has its origin
in a strong phase-mixture as discussed by Luebben et al.
\cite{Luebben2010}. Thus, there exists a Pb substitution region in
the Bi$_{2+z-y}$Pb$_y$Sr$_{2-x-z}$La$_x$CuO$_{6+\delta}$ system
where the superconducting properties might be strongly influenced by
the competition of two different structural phases. However, samples
without Pb and samples with high Pb content ($>$0.3) are then either
dominantly in one phase or the other. This is, why we will discuss
only the series with y=0 and y=0.4. These are shown in FIG.
\ref{FIG2} (a) and (b), respectively. From the averaged T$_C$ for
y=0 (a) and y=0.4 (b), a strictly parabolic-like curve is still not
found. Instead, there are reductions or anomalies of T$_C$.  These
take place for the Pb-free series at La concentration of
x$\approx$0.25$\pm$0.02; 0.43$\pm$0.02 and 0.58$\pm$0.02 formula
units. For y=0.4, an anomaly can be seen at x$\approx$0.4$\pm$0.02
and 0.62$\pm$0.02 formula units and possibly at
x$\approx$0.28$\pm$0.04. But, because of the high variance, the
conclusion of a finer structure within the parabolic curve may yet
be unconvincing. Therefore, we will establish more arguments on
statistical grounds.

\begin{figure}
 \includegraphics[width=0.35\textwidth]{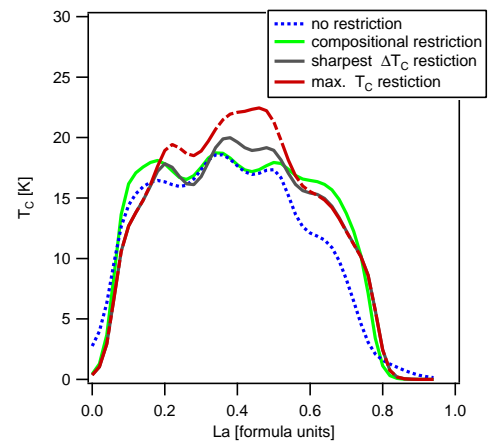}
  \caption{Reducing the number of crystals in order to optimize
  the superconducting properties. Here, the averaged T$_C$ for several
  changes in the ensemble is compared: There is the average of
  the dataset without any restriction (dotted line), the dataset
  with the compositional restriction (solid line), the 'sharpest'
  $\Delta$T$_C$ reduced dataset (dark gray solid line) and the
  'maximum T$_C$' reduced dataset (dashed line). These 'maximum T$_C$'
  reduced dataset and 'sharpest' $\Delta$T$_C$ reduced dataset were
  performed by choosing the samples with the highest T$_C$ or
  sharpest $\Delta$T$_C$ in its lanthanum 'slot' or interval of
  0.02 formula units. One can see that the curves change but show
  always anomalies (suppressions of T$_C$) at about the same lanthanum level.}\label{FIG3}
\end{figure}

In the following analysis, we will prominently regard samples
without Pb. This is merely because there are more data from this
concentration than for y=0.4 and therefore the statistical argument
will weight stronger. The dataset in this y=0 region, filtered by
the compositional restriction, contains 82 samples. One interesting
question is whether the form of the averaged curve of FIG.
\ref{FIG2} (a) is strongly dependent on the chosen ensemble. Whether
the ensemble is representative or not can, of course, never be
determined. We can only apply the following philosophy: The
'compositional restriction' applied before convinced us that the
chemical composition is well defined as the dependence of Bi, Sr,
and Cu versus the lanthanum content is smooth. Variations of the
hardly controllable excess oxygen can be averaged out. Now, more
restrictions can be performed to maybe further optimize the
superconducting properties of the dataset.  These restrictions will
reduce the magnitude of our ensemble and make it possible to test
for changes, i.e. whether the anomalies persist in the T$_C$-curve
or not. For this further refinement, we tested two slightly
different criteria. One is called here the 'sharpest
$\Delta$T$_C$'-criterion while the other one is the 'maximum T$_C$'-
criterion. The idea behind this sharpest $\Delta$T$_C$-criterion is
that one large contribution to the transition width $\Delta$T$_C$
comes from the level of uniformity for the composition on the scale
of the coherence length, whose scale is in the tenth of an Angstrom.
This suggests that from samples, with the same composition measured
by EDX, the sample with a sharper transition width is structurally
better than the sample with a broader transition; It has a better
compositional uniformity below the scale of the typical area
measured by EDX, which is typically on the micron scale. Thereby, it
might be that by this restriction such samples have the best
superconducting properties. A similar argument can be used to
justify the maximum T$_C$-criterion which is taking always that
sample with the highest T$_C$ from the multitude of samples with the
same composition. Technically, first the compositional restriction
is applied. Then, these additional criteria are applied,  by
selecting either that sample with the lowest transition width
$\Delta$T$_C$ or with the highest T$_C$ compared to the other
samples present in the same La-content 'slot'. The slots had a width
of $\Delta$x = 0.02 formula units. Each of these two restrictions
reduces thereby the dataset to 27 samples. After applying these
restrictions, the individual data points are averaged. The result of
this process is shown in FIG. \ref{FIG3}. where the averaged T$_C$
for the sharpest $\Delta$T$_C$-criterion reduced dataset can be
compared to the dataset with only the compositional restriction.
Also shown is the averaged T$_C$ for the maximum
$\Delta$T$_C$-criterion reduced dataset. It is interesting to see
that the curve from samples with the maximum T$_C$ is not the same
as the one with the sharpest $\Delta$T$_C$. Finally also shown is
the averaged T$_C$ for all characterized samples without applying
any restrictions. For comparison, this dataset contains 139 samples.
What can clearly be seen in FIG. \ref{FIG3} is that all the
La-T$_C$-curves change slightly but \textit{always} show anomalies
at almost the same lanthanum concentrations.

\section{Consideration of the hole content}

The methods for determining the exact value of the hole content of
the CuO$_2$-plane in Bi2201 shall be briefly discussed. For, e.g.,
La$_{2-x}$Sr$_x$CuO$_4$ (LSCO), the method to determine the
hole-content is relatively straight forward. Under correct annealing
conditions (see, e.g., \cite{Kanai1997}),
La$_{2-x}$Sr$_x$CuO$_{4+\epsilon}$ has nearly a stoichiometric
amount of oxygen. Therefore, merely determining the value of the
Sr-content x gives the hole content p. In the Bi2201 system, the
situation is more complicated. One reason is that measuring the
quite low concentrations of extra oxygen in tiny single crystals is
an experimentally formidable task. In the publication of Ando et al.
\cite{Ando2000}, the hole content in single crystals was estimated
by comparing the normalized Hall-coefficient of La-Bi2201 to LSCO.
The authors give the dependence  p=0.21-0.13x  for La-Bi2201 in the
substitution range 0.23$<$x$<$0.84. A second method to consider the
hole-content is by the ratio between the Cu-L$_{III}$ white-line and
the Cu-L$_{III}$ charge transfer satellite. These quantities can be
obtained and analyzed in an x-ray absorption spectroscopy (XAS)
experiment. For polycrystalline La-Bi2201 and (Pb,La)-Bi2201, this
method was applied by M. Schneider et al. \cite{Schneider2005}. The
hole scale there differs from the scaling determined by the
comparison with the normalized Hall coefficient by Ando et al.
\cite{Ando2000}. From the XAS measurements, the relation is given as
p=0.24-0.21 x  for La-Bi2201, and p=0.23-0.22x for PbLa-Bi2201. It
should be noted here that the Pb-substituted ceramics had a large
unsystematic variation in the Pb-content. Also for single-crystals,
despite an unusual in-plane polarization dependence
\cite{Mueller2010}, the scaling could be obtained by XAS
\cite{Ariffin2009}. The crystals used there had a more defined Pb
substitution level as the ceramics of Schneider et al.
\cite{Schneider2005}. The scaling of the hole-concentration relative
to the La-content for PbLa-Bi2201 with y$\simeq$0.4 was evaluated in
the single crystals  as p=(0.23$\pm$0.02)-(0.16$\pm$0.05)x.

In the following, we will perform a comparison method to consider
the hole-content. For this, we first will discuss a collection of
T$_C$ data for LSCO from the literature from which we will show that
anomalies of T$_C$ are present in LSCO. Thus, we have the hole
concentrations of anomalies in LSCO and we have the La-content of
anomalies from the series without lead and with lead substitution of
y=0.4 formula units. Eventually, by assigning the hole-value of
anomalies found in LSCO to the La-values of the anomalies found in
La-Bi2201 and PbLa-Bi2201, we can thereby construct a hole scaling
for the number of holes in La-Bi2201 and PbLa-Bi2201. Then, we are
able to discuss the resulting hole scaling in relation to the
hole-scales as given in the literature \cite{Ando2000,
Schneider2005, Ariffin2009} and also discussed above.

\subsection{Anomalies of T$_C$ in LSCO}

\begin{figure*}
  \includegraphics[width=0.75\textwidth]{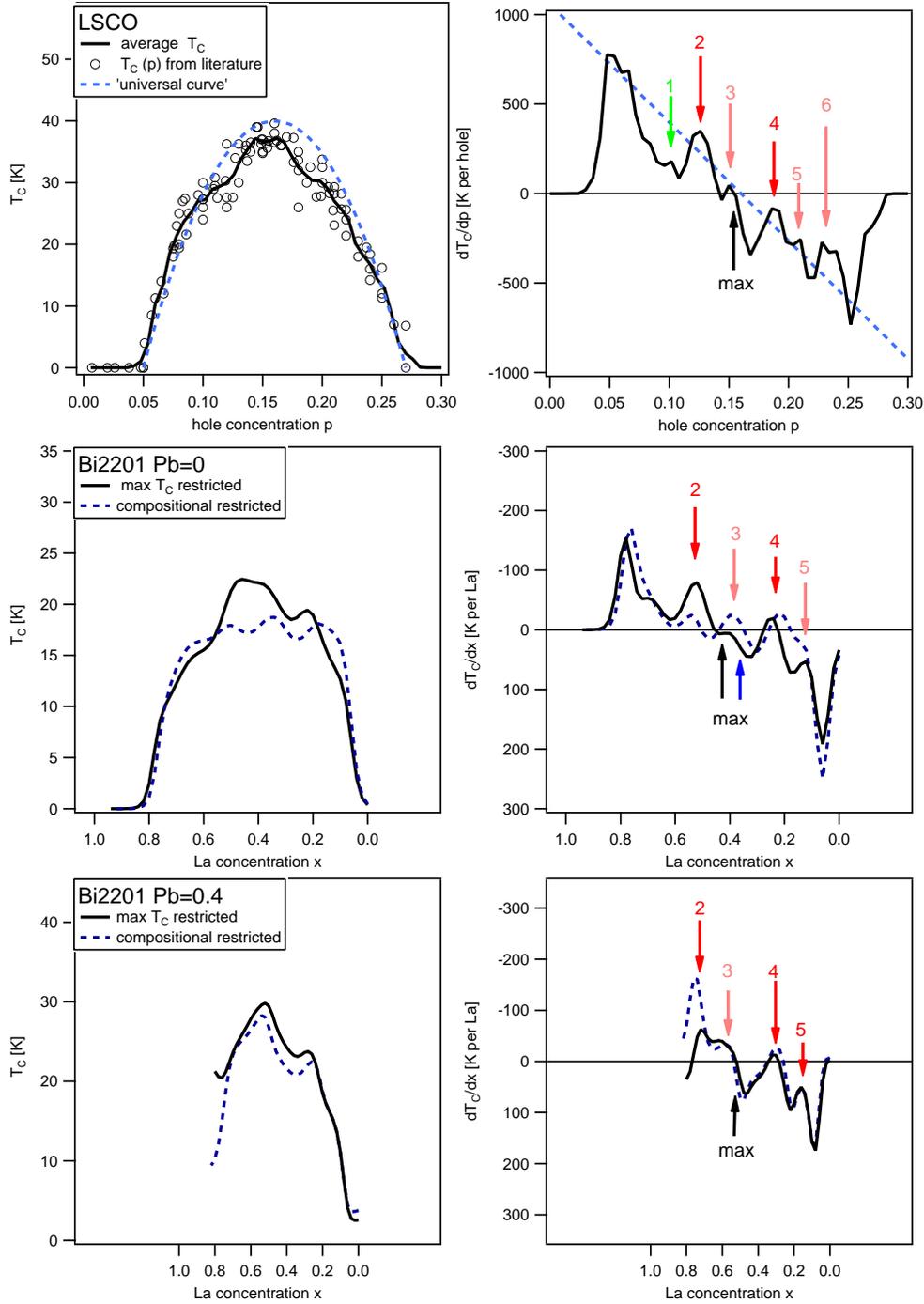}
  \caption{The upper Left panel shows the experimental
  T$_C$ and hole doping values for LSCO as compiled
  from the literature \cite{Suzuki1991, Kimura1992, Nakamura1993,
  Radaelli1994, Shibauchi1994, Fukuzumi1996, Sasagawa2000,
  Hofer2001, Matsuzaki2001, Komiya2005, Yuli2007}. The solid line is
  the averaged curve derived within this work for all the 98 experimental
  data points. The dashed curve represents the 'universal curve',
  computed with the maximum transition temperature T$_C^{max}$=40K. The
  upper right panel shows the first derivative of the average T$_C$ vs
  hole-concentration (p). The anomalies of T$_C$ at hole dopings of
  about x=0.1, 0.125, 0.15, 0.185, 0.21, and 0.235 are marked by
  arrows. The dashed line is the derivative of the 'universal curve'.
  The panels below show the average T$_C$ vs lanthanum concentration
  curves for Bi$_{2+z-y}$Pb$_y$Sr$_{2-x-z}$La$_x$CuO$_{6+\delta}$
  for the Pb contents of y=0 (middle left panel) and y=0.4
  (lower left panel). The derivatives of these averaged T$_C$
  curves are shown respectively on the right panels. The maximum
  T$_C$-restricted average is the solid line and the compositional
  restricted average is the dashed line. The assignment of the anomalies
  is indicated in all right panels by arrows; Also included is the maximum
  position of the T$_C$ curve. Please compare also with TAB. \ref{TAB1}.}\label{FIG4}
\end{figure*}

\begin{table*}
\centering
\begin{tabular}{|c||c|c|c|c|c|c|c|}
  \hline
 &   \multicolumn{6}{c|} {}  &\\
 description & \multicolumn{6}{c|} {label of the maxima of the 1st. derivative} & T$_C^{max}$\\
 & 1 & 2  & 3  & 4  & 5  & 6 & position \\
\hline LSCO x [holes/Cu] $\dag$& 0.98 & 0.126 & 0.15 & 0.188 & 0.21
& 0.228 &
0.156 \\
\hline
magic fractions \cite{Komiya2005} & 3/32  & 1/8  & 5/32  & 3/16  & 7/32  & 15/64 & - \\
as floats:  &  0.094  & 0.125  & 0.156  &  0.188 &  0.219 &  0.234& - \\
\hline
 \hline
 &\multicolumn{6}{c|} {La content x [formula units]}&\\
\hline
La-Bi2201  comp. restr. $\ddag$&- & 0.54 & 0.39  & 0.22  & 0.14    & - & 0.44 \\
La-Bi2201  max. T$_C$ restr. $\ddag$&  - & 0.53  & 0.41  & 0.25  & 0.13  & - & 0.36 \\
(Pb,La)-Bi2201 y=0.4 comp. restr. $\ddag$& -  & 0.75 & 0.58  & 0.30 & 0.16  & - & 0.53 \\
(Pb,La)-Bi2201 y=0.4 max. T$_C$ restr. $\ddag$& -  & 0.72  & 0.58  & 0.31  & 0.16  & - & 0.52 \\
  \hline
\end{tabular}
\caption{\label{TAB1} From top to bottom: The doping of the
anomalies found in LSCO. The 'magic doping fractions'
\cite{Komiya2005} given by the formula p=(2m+1)/2n where m and n are
integers. The positions in La content of the depressions found in
(Pb,La)-Bi2201 for the different averaging methods. On the right
column, the T$_C^{max}$ position of every curve is written. \\
$\dag$ error estimated as $\pm$ 0.005 holes/Cu. \\ $\ddag$ for
individual error estimation please compare with FIG. \ref{FIG5}.}
\end{table*}

T$_C$ values for LSCO were obtained by resistivity measurements and
are taken from \cite{Suzuki1991, Kimura1992, Nakamura1993,
Radaelli1994, Shibauchi1994, Fukuzumi1996, Sasagawa2000, Hofer2001,
Matsuzaki2001, Komiya2005, Yuli2007}. This collection is shown in
the upper left panel of FIG. \ref{FIG4}. Also depicted is the
averaged T$_C$ versus p curve for all the 98 data points. To
generate the curve, four additional data points at p=0.27, 0.28,
0.29, and 0.30 with T$_C$=0 were used. These extra points are used
to improve the visualization and can be justified in that for
p$>$0.26, no superconductivity is detectable. The most remarkable
feature of the T$_C$ vs p curve is that it is also not a simple
parabola for LSCO. This can be seen compared to the 'universal
curve', which is also depicted in the figure and calculated for
T$_C^{max}$ = 40 K. At certain doping concentrations, an anomaly of
T$_C$ occurs. The first derivative of the averaged T$_C$ vs p curve
is shown on the upper right panel of FIG. \ref{FIG4}. These
anomalies can be seen as maxima of $\partial$T$_C$/$\partial$p. Due
to the fact that, in the literature, typically the maximum T$_C$
achievable is given and due to the assumption of a Gaussian
distribution in the averaging algorithm, these singularities are
slightly smeared out. In the upper right panel of FIG. \ref{FIG4},
the anomalies are marked by arrows. They occur at hole dopings of
approximately p=0.098, 0.126, 0.15, 0.188, 0.21  and 0.228 holes per
Cu. T$_C^{max}$ is reached at a doping of about 0.156 holes per Cu.
The error is estimated at about $\pm$0.005 holes per Cu. The
anomalies at p=0.098, 0.15, 0.21 and 0.228 are weaker and the exact
positions more uncertain, while the anomalies at p=0.126 and 0.188
are more pronounced.  The bounce visible in the T$_C$ curve, and
also in the derivative as an upturn around 0.26, is due to the
additionally inserted non-experimental data points described above.
However, it is worth noting that all these anomalies are very well
known, e.g. one anomaly of T$_C$ occurs at hole concentrations of
1/8 \cite{Moodenbaugh1988}. The other anomalies agree with the
'magic doping fractions' found by Komiya et al. \cite{Komiya2005}.

\subsection{Scaling by the anomalies}

As shown earlier, the Bi2201 crystals used here also exhibit
anomalies. These are visible in the T$_C$ vs La curve and can be
used to scale the hole concentrations. This scaling is, of course,
based on the assumption that these anomalies must occur at the same
doping concentrations as those for LSCO. In addition to the
derivative of the average T$_C$ vs hole concentration curve for
LSCO, FIG. \ref{FIG4} shows the derivatives of the average T$_C$ vs
lanthanum concentration curves for
Bi$_{2+z-y}$Pb$_y$Sr$_{2-x-z}$La$_x$CuO$_{6+\delta}$  for the Pb
contents of y=0 and y=0.4. Please note that the La-scale is reversed
compared to the hole-scale of LSCO, as substitution of divalent
strontium by trivalent lanthanum reduces the amount of holes.  The
averages used for the derivatives are the compositionally-restricted
average, which is the data first refined by the compositional
restriction only and then averaged, and the 'maximum
T$_C$-restricted average'. The maximum T$_C$-restricted average is
the data first refined by the compositional restriction, then
refined by the maximum T$_C$ criterion and then averaged. The
maximum T$_C$-restricted average is used because it might better
resemble the features produced by a resistivity measured
T$_C$-curve. This is because, in the case of heterogeneous samples,
by resistivity the path with the highest T$_C$ will be probed. The
maxima of the derivative of the average T$_C$ -curve of Bi2201 were
assigned to these anomalies. This assignment is simply by counting
the 'peaks' in the derivatives from the position of T$_C^{max}$ at
about 16\% hole-doping. In the right side of FIG. \ref{FIG4}, the
positions of the T$_C^{max}$ in the derivatives are marked by blue
or black arrows and labeled 'max'. The assignment of the 'peaks' is
also indicated by arrows in FIG. \ref{FIG4} and also listed in TAB.
1.

\begin{figure*}
  \includegraphics[width=0.75\textwidth]{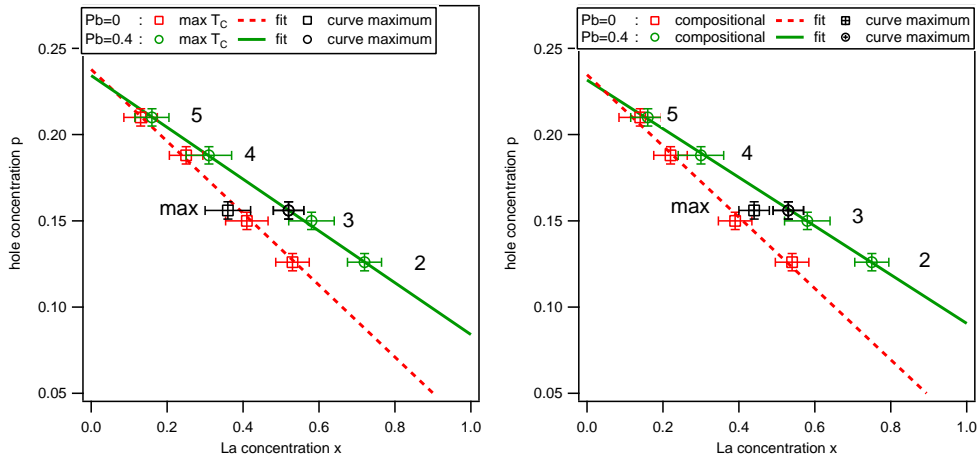}
  \caption{The lanthanum concentration of the maxima in first derivatives
  for both restricted averages vs the hole doping of the maxima found in
  LSCO. Please compare with TAB. \ref{TAB1}. On the left side, the maximum T$_C$
  restricted average is shown for Pb content of y=0 (boxes) and 0.4 (circles).
  On the right side, the same is done for the compositional-restricted average.
  Also, the positions of the maximum transition temperature (T$_C^{max}$) for y=0
  and y=0.4 and both averages vs the T$_C^{max}$ position in LSCO are shown. Also
  included are the linefits for the four restricted averages.}\label{FIG5}
\end{figure*}

FIG. \ref{FIG5} shows the lanthanum concentration of the maxima of
the first derivative for both restricted averages vs the assigned
anomalies. This assignment enables us now to construct a
lanthanum-hole scaling. In agreement with Schneider et al.
\cite{Schneider2005} and Ando et al. \cite{Ando2000}, we assume a
linear behavior for the lanthanum-hole scaling. This assumption can
be justified by a rather small $\chi^2$-value found. A linefit
yields the following relations for the Pb content y=0:
\begin{eqnarray}
p_c&=&(0.23(6) \pm 0.01) - (0.20(9) \pm  0.02) x , \mathrm{and} \nonumber\\
p_m&=&(0.23(9) \pm 0.01) - (0.21(4) \pm  0.01) x . \nonumber
\end{eqnarray}
The first scaling relation, with p$_c$, is the linefit for the
maxima of the first derivative for the compositionally-restricted
average which was the dataset refined by the compositional
restriction only and then averaged. The corresponding averaged curve
is the dashed line in the middle panels of FIG. \ref{FIG4} and the
maxima of this curve are marked by squares on right panel of FIG.
\ref{FIG5}. The second scaling relation, with p$_m$, is the linefit
for the maxima of the first derivative for the maximum
T$_C$-restricted average which was that what was first refined by
the compositional restriction and then -additionally- by the
max-T$_C$ criterion and then averaged. The corresponding averaged
curve is the solid line in the middle panels of FIG. \ref{FIG4} and
the maxima of this curve are marked by squares on left panel of FIG.
\ref{FIG5}.  In both these relations, p is given in holes per Cu and
x is in formula units. Please note the perfect agreement with the
measurements in Schneider et al. \cite{Schneider2005}.  For Pb
content y=0.4, the relations are
\begin{eqnarray}
p_c&=&(0.23(2) \pm 0.01)-(0.14(1) \pm 0.01) x ,
\mathrm{and} \nonumber\\
p_m&=&(0.23(4) \pm 0.01)- (0.14(8) \pm 0.01) x .\nonumber
\end{eqnarray}
Again, the first scaling relation, with p$_c$, is the linefit for
the maxima of the first derivative for the
compositionally-restricted average. The corresponding averaged curve
is the dashed line in the lower panels of FIG. \ref{FIG4} and the
maxima of this curve are by circles on right panel of FIG.
\ref{FIG5}.  The second scaling relation, with p$_m$, is the linefit
for the maxima of the first derivative for the maximum
T$_C$-restricted average. The averaged curve is the solid line in
the lower panels of FIG. \ref{FIG4} and the maxima of this curve are
by circles on left panel of FIG. \ref{FIG5}. Here, the measurements
do not concur as well with Schneider et al. \cite{Schneider2005}. As
already mentioned, the discrepancy might have its origin in the
unsystematic variation of Pb content there. The agreement with the
more Pb-homogenous single crystals used by Ariffin et al.
\cite{Ariffin2009} is good.

\begin{figure*}
  \includegraphics[width=0.8\textwidth]{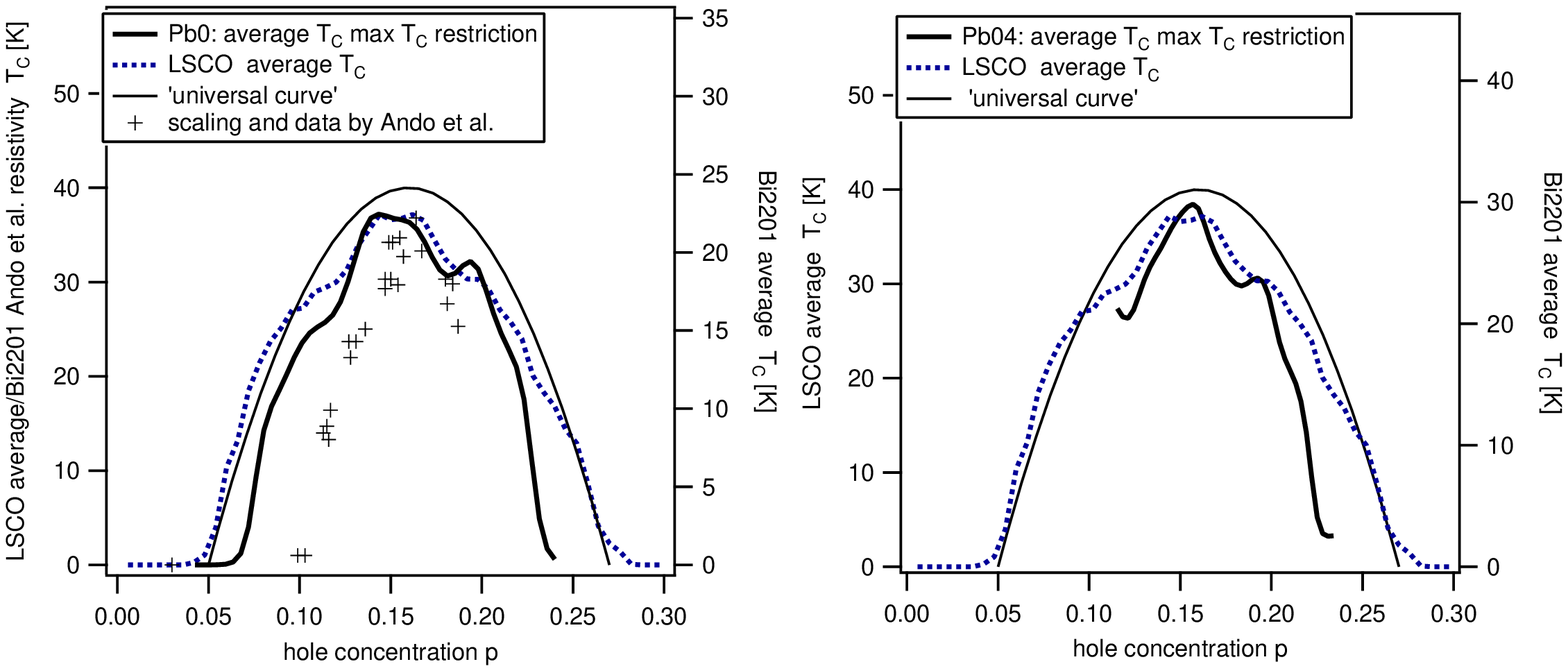}
  \caption{Lanthanum hole scaling by the fit from the maximum T$_C$ average (pm).
  Left for Bi$_{2+z}$Sr${2-x-z}$La$_x$CuO$_{6+\delta}$. There also included is the comparison with
  the scaling and data from Ando et al. \cite{Ando2000}. Right for
  Bi$_{2+z-y}$Pb$_y$Sr$_{2-x-z}$La$_x$CuO$_{6+\delta}$
  with Pb content y=0.4. For both graphs the T$_C$-curve for LSCO and the
  'universal curve' are also shown.}\label{FIG6}
\end{figure*}

In FIG. \ref{FIG6}, finally, the T$_C$ vs hole concentration curves
achieved by this scaling are shown. The scaling used is the fit from
the max. T$_C$ average (pm). The left side of FIG. 6 shows the
scaling of the max. T$_C$ average of
Bi$_{2+z-y}$Pb$_y$Sr$_{2-x-z}$La$_x$CuO$_{6+\delta}$ for Pb content
y=0 whereas the right side shows the same for the y=0.4 series. Also
included for Pb content y=0 is the comparison with the scaling and
data of Ando et al. \cite{Ando2000}. For both Pb concentrations, the
averaged T$_C$-curve for LSCO and the 'universal curve' are also
shown. With respect to LSCO and the 'universal curve', the curve for
Bi$_{2+z-y}$Pb$_y$Sr$_{2-x-z}$La$_x$CuO$_{6+\delta}$  with Pb
content y=0  shows a faster drop at the underdoped and overdoped
sides. The same trend is obvious for the y=0.4 result of FIG.
\ref{FIG6} although the final conclusion may suffer from the limited
data set.

\section{Discussion}

In this publication, we showed that the superconducting transition
temperature for
Bi$_{2+z-y}$Pb$_y$\-Sr$_{2-x-z}$La$_x$CuO$_{6+\delta}$ with y=0.4
and Bi$_{2+z}$Sr$_{2-x-z}$\-La$_x$CuO$_{6+\delta}$  exhibit
anomalies at certain lanthanum concentrations. A hole scaling was
achieved by assigning the anomalies visible in the T$_C$ vs
lanthanum graph to certain dopings. This assignment gives  the same
hole doping vs lanthanum scaling derived with x-ray absorption
spectroscopy by Schneider et al. \cite{Schneider2005} and Ariffin et
al. \cite{Ariffin2009}. It is suggested that the difference in the
scaling-relation for Bi2201 derived here and by Schneider et al.
\cite{Schneider2005} compared to the scaling derived in
\cite{Ando2000} is due to a difference in crystal growth, the method
of determining the chemical composition and the differences in
methods for obtaining the hole-concentrations. In fact, analyzing
the T$_C$(p) data of Ando et al. \cite{Ando2000}, as reproduced on
the left side of FIG. 6, there is a change in curvature at around
0.14 to 0.15 holes per Cu which we attribute to 1/8. Rescaling the
hole-doping vs lanthanum equation of \cite{Ando2000} with this
knowledge and the knowledge that T$_C^{max}$ is at around a
lanthanum level of 0.4 does not yield our hole-doping vs. lanthanum
scaling. Unfortunately, we found it impossible to estimate possible
compositional differences other than the lanthanum between our
crystals and the crystals of Ando et al. \cite{Ando2000}.
Differences might be, for example, in the Bi-Sr exchange. However,
as here the same scaling was achieved by either a direct
quantitative method \cite{Ariffin2009, Schneider2005} and
additionally by a comparison method, the hole-concentration curve
presented here is validated, at least for crystals with similar
chemical composition and this composition obtained by a similar EDX
processing.

In the course of this manuscript, we derived a hole-doping scale by
an assignment of the La-value of the anomalies to the corresponding
hole-values found for LSCO. Using the hole-scaling of Schneider et
al. \cite{Schneider2005} and Ariffin et al. \cite{Ariffin2009}, we
can turn the line of argumentation and then say that two kinds of
Bi-cuprates show anomalies at very much the same hole-doping values
as they are in LSCO. These extracted dopings of the anomalies agree
with the 'magic doping fractions' found in LSCO by Komiya et al.
\cite{Komiya2005}. There, the anomalies were measured as maxima in
the normal-state (or pseudogap-state) resistivity. Here, we showed
that with a large enough amount of data, the anomalies can be seen
directly in the T$_C$(p) curve - without using high temperature
signatures like it was done by Komiya et al. \cite{Komiya2005}. But,
it is worth to mention that our resistivity measurements of the
ab-plane also confirm a more insulating behavior at the anomalies.
It is suggested that the identification of the anomalies by its high
temperature signatures is possible due to the intimate relation
between the anomalies and the pseudogap state. However, this issue
is beyond the scope of this publications and will be addressed
elsewhere.

By the last paragraph, we hope to have made clear that, in our view,
the existence of a multitude of anomalies of T$_C$ is proved for
two-and-a-half cuprate systems: LSCO and two variants of Bi2201. The
'half' is related to the already mentioned fact that, with large
enough amount of Pb, the Bi2201 crystals are dominantly in a
different structural phase (see, e.g., \cite{Luebben2010}). We can
next discuss the question whether the cause of these anomalies is a
structural effect. For this, let us look at the following three
facts: (1) The CuO$_2$-plane in
Bi$_{2+z-y}$Pb$_y$Sr$_{2-x-z}$La$_x$CuO$_{6+\delta}$  with y=0.4
exists in a different structural environment than in
Bi$_{2+z}$Sr$_{2-x-z}$La$_x$CuO$_{6+\delta}$. (2) The structural
environment of La$_{2-x}$Sr$_x$CuO$_4$ is quite different from both
of the Bi-cuprates. (3) The amounts of dopant-atoms for reaching
optimal doping are very much different in all the three systems
(La-Bi2201 (Pb,La)-Bi2201 and LSCO). Following this line of
argumentation, structural effects can play no major role here. Also,
for La-Bi2201 and PbLa-Bi2201, by the observation of the anomalies
at different La-contents but the same hole-dopings, variations in
the hole-doping due to the variation of the poorly controlled excess
oxygen can be ruled out as a cause. In fact, the effect of the
oxygen was already removed by our statistical Ansatz and the
variation of the ensemble shown here. This statistical Ansatz seems
generally beneficial - also for LSCO, where the hole-doping is
better controllable than in the Bi-cuprates. The derived average
T$_C$ decreases at an anomaly, but it seems to us that the tendency
of having a somewhat different T$_C$ increases, a behavior which we
can also slightly detect in the standard deviation of our averaging
(see FIG.2). It is also an interesting question, whether this
tendency of having somewhat different T$_C$'s is caused somehow by
the mechanism of superconductivity in the CuO$_2$-plane itself or an
indirect effect due to another parameter not well controlled or a
combination of both. For the influence of a not well controlled
parameter, an example is the 60K plateau in YBCO, the 1/8 anomaly
\cite{Tallon1997}, which can be reduced by ordering of the oxygen
within the CuO-chains of the charge reservoir (please compare with
\cite{Poulsen1991, Tallon1997, Segawa2001}). Finally, another
interesting question is why some anomalies, like 1/8 and maybe 5/32,
seem to be more pronounced than others. A quantitative measure of
this might be the different heights of the 'peaks' appearing in the
derivative of the T$_C$ curve (FIG. \ref{FIG4} right column).

Let us now discuss possible mechanisms causing these anomalies. For
this, we review some experimental findings, mostly related to the
1/8 anomaly, and some theoretical concepts:

It is known that neutron scattering detects fourfold superlattice
peaks indicating a spatial spin-ordering \cite{Cheong1992,
Tranquada1995, Yamada1997}. X-ray-diffraction \cite{Zimmermann1998}
and resonant soft x-ray scattering \cite{Abbamonte2005} also
indicates charge ordering. A study for LBCO by angular resolved
photoemission and scanning tunneling spectroscopy \cite{Valla2006}
showed that at 1/8 the d-like single particle gap shows a maximum.
In La$_{1.6-x}$Nd$_{0.4}$Sr$_x$CuO$_4$, neutron scattering revealed
\cite{Tranquada1995} the superlattice peaks being commensurate with
the underlying lattice. Assuming that this commensurability for
La$_{1.6-x}$Nd$_{0.4}$Sr$_x$CuO$_4$ is a typical sign at 1/8. In
this case, one idea might be that at 1/8 are static (or
commensurate) stripes (for a review see, e.g.,
\cite{Orenstein2000}). Thus, there is a formation of hole-rich,
quasi-one dimensional 'rivers of charge' separated by
antiferromagnetic domains - an idea which has its roots in the
mean-field results of Zaanen and Gunnerson \cite{Zaanen1989}. In
order to have this unidirectional order consistent with neutron
scattering fourfold symmetry, there must be two domains rotated 90
degree to each other, most preferentially there is a stacking with
the stripes rotated by 90 degree for each neighboring
CuO$_2$-planes. However, unlike dynamical (or incommensurate)
stripes, Emery et al. \cite{Emery1997} stated that commensurability
leads to a charge gap and insulating behavior which competes with
superconductivity. There can be a commensurability effect within a
stripe, which tends to pin when 2k$_F$=2p/m, where m is an integer
and k$_F$ the Fermi-wavevector. There is also a commensurability
effect between stripes. In the most conventional form, one can think
of half-filled commensurate stripes which occur at hole-dopings
p=1/2n. Here, n is an integer representing the spacing between the
stripes.

Motivated by a pattern found in the power spectral density of
topological data of scanning tunneling microscopy (see
\cite{Fischer2007} and references therein), another relevant model
might be the pair-density wave state (PDW-1) as described by Chen et
al. \cite{Chen2004}. When looking at the mean field result of an
SO(5) model with extended interactions, it is found that there are
insulating phases at fractional filling factors where Cooper pairs
of the holes form a lattice which is usually commensurate with the
underlying lattice. Thus, there is a phenomenological state where a
non-vanishing superconducting order can coexist with a charge
density wave order. These insulating PDW-1 states are a consequence
of strong pairing and low superfluid density. By assuming a 'Law of
Corresponding States', it can be predicted that these insulating
PDW-1 states appear in the background of superconducting states at
rational fillings of p=(2m+1)/2n \cite{Komiya2005}. Here, p is the
hole-doping and m and n are integers. Let us point out that, unlike
the static stripe scenario above, the order in the CuO$_2$-plane is
two-dimensional. However, the existence of these PDW-1 states is
dependent on the strength of an interaction parameter. There are
therefore materials which are believed to show no anomalies, one
example being Bi$_{2+z}$Sr$_{2-z}$CaCu$_2$O$_{8+\delta}$ (Bi2212)
\cite{Chen2004}. By the construction principle, there is also a
hierarchy of these PDW-1 states, where 1/8 is the most pronounced
and most common anomaly followed equally by the more uncommon 3/16
and 1/16 anomalies and so on (see \cite{Komiya2005}).

Another concept is that there is Wigner-crystallization of holes in
an antiferromagnetic background as Kim and Hor \cite{KimHor2001,
KimHor2006} deduced by analyzing far-infrared reflectivity
measurements. At the hole-values of the anomalies and below the
crystallization temperature an insulating hole-lattice is formed
which is pinned to the underlying lattice, and therefore
commensurate with it. Extra carriers (holes) occupy the interstitial
sites and can 'ride' the Wigner lattice forming an 'interstitial
band'. These extra carriers form spin-singlet Cooper pairs because
they experience the negative dielectric screening in the frequency
region between the Goldstone mode (in this situation, the phononic
mode of the Wigner crystal) and the plasma frequency of the
hole-lattice. Unfortunately, these ideas seem not to have triggered
much further theoretical considerations. One may find it interesting
to compare the picture of an 'interstitial band' with calculations
on the Hubbard-Wigner model, for example exact diagonalization
results for a finite size lattice \cite{Fratini2009} or results by
single-site dynamical mean field theory \cite{Amaricci2010}. It is
also interesting that, in a (true) Wigner crystal, path-integral
Monte Carlo calculations \cite{Candido2001} do show strong
attractive interaction between point defects. We would like to
emphasize that, of course, in these correlated systems a
crystallized hole-lattice with point defects in an antiferromagnetic
background has some more constraints as just those given by the
confinement on a lattice, an onsite-exchange interaction and
long-range Coulomb interactions. That is because the Zhang-Rice
singlet picture breaks down above p=0.25 holes per Cu when nearest
neighbor singlets would then have to share a mutual oxygen (compare
with \cite{Roehler2005, Peets2009}). However, for the doping values
of the anomalies, Kim and Hor \cite{KimHor2006} expect them at p =
m/n$^2$ with nonzero integers m and n where m$<$n$^2$.

For LBCO, it seems to be that the onset of charge and spin order
leads to the suppression of Josephson coupling between neighboring
CuO$_2$-planes \cite{Li2007, Tranquada2008}. In underdoped LSCO, it
was shown that, when applying a moderate magnetic field
H$\parallel$c, the interlayer coupling is suppressed albeit the
in-plane superconducting properties remain intact
\cite{Schafgans2010}.  The model of the 'striped superconductor'
\cite{Berg2007, Berg2009} seems to be especially suitable to explain
these experimental findings as it leads easily to the so-called
'dynamical layer decoupling' because the average c-axis
Josephson-coupling vanishes. In this model, there is a time reversal
invariant state in which the superconducting order parameter has a
finite wave vector and changes sign over half a period. This state
is confusingly also called a pair-density wave (PDW-2) but with a
different meaning than for the one of Chen et al. \cite{Chen2004}.
While for PDW-1 the (non-vanishing) superconducting order parameter
coexists with the order parameter of the charge density wave, for
PDW-2 the average value of the superconducting order parameter
vanishes and there is a charge density wave order induced with half
the period of the modulation of the superconducting order parameter.
Going now from the phenomenological picture to a microscopical view
\cite{Berg2009}, there is the picture of a superconducting striped
state where neighboring stripes are antiphase in terms of the
superconducting order parameter. In a layered system and with
perfect stripe order within the planes, the antiphase
superconducting order is able to greatly reduce the interplane
Josephson couplings without the suppression of the antinodal gap.
While the layer decoupling effect can be expected to be most
pronounced where perfect stripe order is near 1/8,  it also occurs
for some range about 1/8. Another major difference of this model
compared to the others discussed above is that before the anomalies
were commensurate with the underlying lattice. In this PDW-2 model,
there is no such commensurability and it is not as important as it
is for the idea of static stripes, which was the first model
reviewed here. A problem with static stripes is that, on the one
hand, neutron scattering gives for 1/8 lower width of the
superlattice peaks and there seems to be a saturation in the doping
dependency of the superlattice peak's periodicity at 1/8
\cite{Yamada1998}, but, on the other hand, in most of the nominal
1/8 samples of various HTSC's, the fourfold superlattice peaks are
not interpreted to be originating from static stripes but dynamical
ones. Often, within the view of stripes, the term 'quasi-static
stripes' is used for this lack of perfect commensurability. In this
view of the PDW-2 state, it might be a question whether
commensurability is playing a major role in the 1/8 problem.

In order to discuss possible causes for not only 1/8 but for all the
anomalies, we may say that all the theoretical models discussed
above for 1/8 might be generally relevant but the weighting of some
details may favor one or the other model. One criterion might be
that for models favoring a unidirectional order, like the PDW-2
state of Berg et al. \cite{Berg2007, Berg2009}, it might be more
challenging to explain the other anomalies found - i.e. others than
at 1/8. Our derived values for the anomalies were quite similar to
those of Komiya et al. \cite{Komiya2005}, which suggests a
two-dimensional order of some kind. On the other hand, this PDW-2
model explains the dynamical layer decoupling quite well, which is
where the PDW-1 state of Chen et al. \cite{Chen2004} seems to have a
problem. A coexistence of charge-order and superconducting-order
would generally reduce T$_C$ and not leave the in-plane
superconducting properties intact. However, this PDW-1 model
reproduces quite well the doping values of the anomalies. As the
deviation of the doping values of the anomalies with p=(2m+1)/2n is
essentially by a two-dimensional geometrical construction scheme, we
have some hopes that something similar may also apply to a model
much like the single hole Wigner crystallization model of Kim and
Hor \cite{KimHor2001, KimHor2006}.

In summary, we showed that there are anomalies of T$_C$ in La-Bi2201
and (Pb,La)-Bi2201 at certain hole-dopings. These hole-dopings agree
well with the extracted 'magic doping fractions' found in LSCO by
Komiya et al. \cite{Komiya2005}. In addition, from a collection of
data taken from the literature, we showed for LSCO that with a large
enough amount of data the anomalies can also be seen directly in the
T$_C$(p) curve. This, together with the knowledge of the 1/8 anomaly
in LBCO and YBCO, allows us to point to the strong possibility that
\textit{all these anomalies are generic for the hole-doped
high-temperature superconductors}. The need to include these generic
anomalies of T$_C$ should help to privilege, expand or create
theories which eventually will explain the occurrence of
superconductivity in these materials.

\subsection{Acknowledgement} We gratefully thank R. M\"uller, J. R\"ohler,
E. Fradkin and J.W. Allen for valuable discussions during
preparation of this publication. We thank M. Mioduszewski for
critical reading of the manuscript. L.D. thanks J.W. Allen for his
kind hospitality during finishing of this manuscript. We thank D.
Kaiser for helping us greatly with the digitalization of the
characterization list.

\begin{widetext}
\appendix
\section{Gaussian slotting algorithm}
\label{gaussianslotting}

Here, we briefly describe the used averaging algorithm. The
advantage of the algorithm is that non equispaced data can be used
and that the data is weighted by its error. As usual, we write the
chemical composition of La as $x$ and the chemical composition of Pb
is $y$. For the La-Pb phase diagram there are existing the
non-equispaced continuous sampled dataset of the form
$T_C^j(x^j,y^j)$, where $j=0,1,\cdots (J-1)$ is the indicae of the
$J$ measured samples. As a result of the EDX measurement, we have a
standard error for the composition, $\sigma_x^j$ and $\sigma_y^j$.
It contains the systematic error of EDX due to the fitting process
and the statistical error from probing multiple areas on the
samples' surface. From the multitude of these non-equispaced and
continuous $T_C^j(x^j,y^j)$, we want to construct an equispaced
discrete averaging function, here formally written as
$\overline{T_C}(p\Delta x,q\Delta y)$. This can be done by weighting
each $T_C^j(x^j,y^j)$ with its standard error into each 'slot'
$(p\Delta x,q\Delta y)$. Here, $p=0,1,\cdots (P-1)$ and
$q=0,1,\cdots (Q-1)$ are the indicae of the slot. The averaging
reads
\begin{equation}
\overline{T_C}(p\Delta x,q\Delta y)=  \frac{1}{Z(p\Delta x,q\Delta
y)} \int\limits_{p\Delta x-\Delta x/2}^{p\Delta x+\Delta x/2} dx'
\int\limits_{q\Delta y-\Delta y/2}^{q\Delta y+\Delta y/2} dy'
\sum\limits_{j=0}^{J-1}
\frac{T_C^j(x^j,y^j)}{2\pi\sigma_x^j\sigma_y^j}\,\times \exp
\left\{-\frac{(x'-x^j)^2}{2(\sigma_x^j)^2}-\frac{(y'-y^j)^2}{2(\sigma_y^j)^2}\right\}.
\nonumber
\end{equation}
Here $Z$ is the probability distribution and given as
\begin{equation}
Z(p\Delta x,q\Delta y)= \int\limits_{p\Delta x-\Delta x/2}^{p\Delta
x+\Delta x/2} dx' \int\limits_{q\Delta y-\Delta y/2}^{q\Delta
y+\Delta y/2} dy' \sum\limits_{i=0}^{J-1}
\frac{1}{2\pi\sigma_x^i\sigma_y^i}\,
\exp\left\{-\frac{(x'-x^i)^2}{2(\sigma_x^i)^2}-\frac{(y'-y^i)^2}{2(\sigma_y^i)^2}
\right\} \, . \nonumber
\end{equation}
It is easy to see that the normalization makes sense:
\begin{equation}
\sum\limits_{p,q=-\infty}^{\infty} Z(p\Delta x,q\Delta y)= J \approx
\sum\limits_{p,q=0}^{(P-1)(Q-1)} Z(p\Delta x,q\Delta y). \nonumber
\end{equation}
The largest error in this approximation comes from boundary effects
of samples with $x-\sigma_x<0$ or $y-\sigma_y<0$. Without loosing to
much accuracy, for $\Delta x, \Delta y \ll \sigma_x, \sigma_y$ the
integrals above can be reduced to
\begin{equation}
\overline{T_C}(p\Delta x,q\Delta y)\approx \frac{\Delta x \, \Delta
y}{Z(p\Delta x,q\Delta y)} \sum\limits_{j=0}^{J-1}
\frac{T_C^j(x^j,y^j)}{2\pi\sigma_x^j\sigma_y^j}\,\times \exp
\left\{-\frac{(p\Delta x-x^j)^2}{2(\sigma_x^j)^2}-\frac{(q\Delta
y-y^j)^2}{2(\sigma_y^j)^2}\right\}\, .  \nonumber
\end{equation}
The probability reads then after reduction
\begin{equation}
Z(p\Delta x,q\Delta y)\approx \Delta x \, \Delta y
\sum\limits_{i=0}^{J-1} \frac{1}{2\pi\sigma_x^i\sigma_y^i}\,
\exp\left\{-\frac{(p\Delta x-x^i)^2}{2(\sigma_x^i)^2}-\frac{(q\Delta
y-y^i)^2}{2(\sigma_y^i)^2} \right\} \, . \nonumber
\end{equation}
These both approximations above were used for computing the averaged
T$_C$-function. This averaging function is expected to be
asymptotical unbiased as long as the estimated function, i.e. the
true T$_C$ function by hypothetically measuring an infinite fine
doping series of perfect samples, is analytically and harmless. It
is clear, that the variance of this averaging increases for slots
with a low probability, i.e. with 'no samples in the slot'. For
visualization of $\overline{T_C}(p\Delta x,q\Delta y)$ it is
therefore good to define a cutoff value. Below this value the
function  is not plotted. Therefore, it should be that
\begin{equation}
Z(p\Delta x,q\Delta y) \geq Z^{\mathrm{cutoff}}:=\frac{1}{2PQJ}.
\nonumber
\end{equation}
From the normalization given above it can be seen that this means
that half a samples probability has to be located in the slot.

\end{widetext}

\bibliography{bibliography_endversion}

\end{document}